\documentstyle[preprint,aps,prb,epsfig]{revtex}
\def\Frac#1#2{\frac{\displaystyle #1}{\displaystyle #2}} 
 
\begin{document}
 
\newcommand{\beq}{\begin{equation}}
\newcommand{\eeq}{\end{equation}}
\draft

\bibliographystyle{unsrt} 

\title
{\bf A FUNCTIONAL APPROACH TO NUCLEAR \\
    ELECTROMAGNETIC RESPONSE FUNCTIONS}
\author{R. Cenni, F. Conte and \underline{P. Saracco}}
\address{INFN - sez. Genova - Via Dodecaneso, 33 - I16146 Genova (Italy)}

\maketitle

\begin{abstract}
The separated electromagnetic responses $R_L(q,\omega)$ and 
$R_T(q,\omega)$ for inclusive electron scattering off nuclei 
are studied within a functional scheme. 
\end{abstract}
 

 
\narrowtext


\section{Introduction}

During the '80s experimentalists were able to separate out of the inclusive
electron-nucleus cross-section the charge and magnetic response functions via
a Rosenbluth plot 
whose validity is grounded on two assumptions: the goodness of the
one-photon-exchange approximation and the negligibility of the Coulomb
distortion of the electrons.

Since then a large amount of theoretical speculations was devoted to 
the unexpected behaviour of $R_L(q,\omega)$ and $R_T(q,\omega)$:
in particular $R_L(q,\omega)$ seems largely quenched with
respect to the independent particle model, while $R_T(q,\omega)$ is
somehow enhanced, a fact this last which is not so often emphasized in
the literature.

The exact amount of the quenching of $R_L(q,\omega)$ (and of the correlated
enhancement of $R_T(q,\omega)$) seems to depend upon the target nucleus,
indicating larger effects in medium-heavy nuclei.
Recently Jourdan\cite{Jourdan-1995,Jourdan-1996a,Jourdan-1996b} 
addressed two questions:
first of all about the internal coherence of the known experimental data and 
secondly about their interpretation in terms of the Coulomb sum
rule. 

He re-examined the whole set of the available experimental data for inclusive
electron scattering ("world data") and he was able to perform a 
cleaner and more reliable Rosenbluth separation of $R_L(q,\omega)$ and
$R_T(q,\omega)$ than those based on a single
experimental set. This essentially for two reasons: firstly a better
treatment of the Coulomb distortion and secondly because using the world
data enables one to perform the Rosenbluth separation spanning a larger
interval in 
$\varepsilon\equiv\left(1+2|{\bf q}|^2/Q^2\tan^2 \theta/2\right)^{-1}$,
the kinematical variable for the plot, than the one spanned by a single
experimental set of data.
The net result of this procedure is twofold: first of all one obtains 
a set of "corrected" or hopefully more reliable experimental response functions
and secondly the apparent failure of the Coulomb sum rule is washed out 
at a sufficiently higher $q$
(for instance $q\sim 600 MeV/c$).

Still the results for the response functions display in the intermediate
momentum region some non trivial behaviours, $R_L$ being anyway largely 
quenched with respect to FFG predictions [in an A-dependent way] and $R_T$
being somehow enhanced.
So the question of the behaviour of the separate e.m. response functions
remains in our opinion largely not
understood.
Moreover since some doubts remain even on the procedure followed by
Jourdan - not in principle, but because at larger $\varepsilon$ available
data are older and few and because these data are mostly responsible for
the corrections - we will continue to present our calculations against
both Jourdan data and the original experimental data, when both are
available at the given momentum. We emphasize that it would be highly
desirable
that future experiments (at TJNAF?) would unambiguously solve the 
experimental problem.

\section{The theoretical situation and the functional approach.}

A wide set of different theoretical calculations based on a variety
of dynamical models yields a more or less pronounced depletion of
the QEP in the longitudinal 
channel\cite{Fantoni-1987,Co-1988,Barbaro-1996,VVder-1996}. It is beyond
the purposes of this contribution to provide a detailed comparison
between them. Most of them are anyway unable to provide a good description
of $R_T$ also.

Some years ago we proposed a theoretical scheme based on the application
of the SPA to a (finite) generating functional describing a relativistic
system of nucleons and pions\cite{AlCeMoSa-87}. 
SPA was applied after explicit integration of the nucleonic degrees of freedom:
in this way the presence of a nuclear medium is treated in principle 
correctly. To make such a scheme useful in practice we needed to restrict
it to a nonrelativistic potential theory including the interaction only
in the particle-hole channel with the quantum numbers of a 
pion\cite{AlCeMoSa-90}, later on we included also the interaction 
in the channels with the quantum numbers of 
$\rho$ and $\omega$ mesons\cite{CeSa-94} to describe $R_L(q,\omega)$ alone. 
Finally we considered the contributions coming from
the excitation of a $\Delta_{33}$ resonance, which are known to
be necessary when $R_T(q,\omega)$ is under consideration\cite{CeSa-95} and
we extend now the interaction to cover all the ph channels with
$T=0,1$ and scalar (S), spin-longitudinal (L) and spin-transverse (T).

We do not re-propose here the theoretical derivation of our approach,
since it can found in the literature\cite{CeSa-94,CeSa-95},
but we must spend few words about its dynamical content. To make practically
manageable the calculation we need an effective interaction given in each
spin-isospin channel as a function of the transferred momentum only, including
obviously NN, N$\Delta$ and $\Delta\Delta$ transitions. Its form 
has been chosen to resemble a very traditional parametrization of the 
effective interaction in the $\pi$ and $\rho$ channels, namely
\beq
V_{L,T}^{T=1}(q)=\Frac{f_{\pi NN}^2}{m_\pi^2}
\left\{g^\prime(q)-C_{L,T}\Frac{q^2}{q^2+m_{\pi,\rho}^2}\right\}
v^2_{L,T}(q^2)
\eeq
with $C_L=1$ and $C_T\simeq 2.3$. The Landau parameter $g^\prime$ is usually
assumed to be constant and set around $0.6\div 0.7$;
we allow instead  $g^\prime$ to be momentum-dependent. 
In mesonic scheme it is convenient to
explicitly use meson-exchanges only for the medium/long range
part of the interaction. The remaining part can be either parametrized with
constants or with slowly varying functions of the momentum, or, alternatively
making use of two-body correlation functions. These last can be either 
explicitly introduced or one can simply admit that
\beq
\lim_{q\to 0}g^\prime_{L,T}(q)=g^\prime\,,\qquad\qquad
\lim_{q\to\infty}g^\prime_{L,T}(q)=C_{L,T}
\label{eq:cor}
\eeq
in such a way to properly cut the high-momentum components of the meson exchange
potentials. All these techniques are conceptually equivalent, provided one
keeps in mind the physical difference between the vertex cutoff $v^2_{L,T}(q^2)$
and the many-body correlation length implicit in eqs.~(\ref{eq:cor}). By
choosing arbitrarily the functional form
\beq
g^\prime_{L,T}(q)=C_{L,T}+\left(g^\prime-C_{L,T}\right)
\left[\Frac{q^2_{c-L,T}}{q^2+q^2_{c-L,T}}\right]^2\,,
\eeq
we are led to determine the momentum dependence of $g^\prime_{L,T}(q)$ simply
through the choice of two parameters $q_{c-L,T}$, which are expected in the
range
$ m_{\pi,\rho}<q_{c-L,T}<\Lambda_{L,T}$,
owing to the previous discussion; the $\Lambda$s are the usual vertex cutoffs:
$v_c(q^2)=(\Lambda_c^2-m_c^2)/(\Lambda_c^2+q^2)$.
In the $\omega$'s channels we simply choose an effective interaction of the 
form
\beq
V_\omega^{S=T}(q)=-\Frac{f_{\pi NN}^2}{m_\pi^2}
C_\omega\Frac{q^2}{q^2+m_{\omega}^2}
v^2_{\omega}(q^2)\,,\qquad
V_\omega^{S=0}(q)=\Frac{f_{\pi NN}^2}{m_\pi^2}
C_{\omega_S}\Frac{q^2}{q^2+m_{\omega}^2}
v^2_{\omega_L}(q^2)
\eeq
and an analogous form is chosen for the $S=0$ component of the $\rho$. The
corresponding potentials when $N\Delta$ and/or 
$\Delta\Delta$ transitions can occur are obtained by 
replacing $f_{\pi NN}$ with $f_{\pi N\Delta}$ or with $f_{\pi\Delta\Delta}$
and modifying the vertex cutoffs. The complete set of parameters we are
employing are reported in the appendix. Obviously all these potential
are multiplied by the proper combination of spin-isospin matrices for
the given channel.

Some comments are here in order regarding the procedure we have followed
to determine the parameters in the effective interaction deferring a full
discussion to a
next-to-come paper\cite{CeSa-96a}. The most part of the coupling 
parameters and all the vertex cutoffs are chosen accordingly to a "democratic
principle" looking to the current literature: 
essentially they are set to values coming
either from mesonic realistic potentials (like, e.g., the Bonn potential) or, when
the former possibility is precluded, by using quark-model indications.
The two many-body cutoffs have been tentatively set to 800~MeV
and 1300~MeV in the $\pi$ and $\rho$ channels respectively.
The mass of the transverse $\rho$ has been set to $600~MeV$, a value which
should keep memory of the attraction felt by the $\rho$ inside the
nuclear medium both when living as a $\rho$ and as a couple
of pions. The value of $g^\prime(0)$ has been set to $.35$; these values
completely determine the potential both in the $\pi$ and in the $\rho$ 
channels and have been chosen to reproduce as far as possible the known
effective interactions in the low/intermediate $q$-range and allowing a faster
decrease of the high-$q$ tails in order to describe two-body correlations.
The low value of $g^\prime$ should not worry about the possibility of pion
condensation, since this can happen at typical values of $q\sim 2k_F$, where
$g^\prime_L(q)\simeq 0.75$.

The only value we took as a practically free parameter is the coupling for
the scalar component of the $\omega$. We fixed it naively to
a value around 0.15 by simply fitting the longitudinal response function
for $^{12}C$ at $q=300~MeV$. This so small value is obviously
an effective one and could emerge, for instance, as a parametrization
to the ladder series of "bare" scalar $\omega$'s.

Clearly both this point and the determination of the effective $\rho$ mass 
require a further microscopic investigations. We are confident
that the chosen values are not so far from reality since they provide
a qualitative description of nucleon (and $\Delta$) self-energies.

\begin{figure}
\begin{center}
\mbox{
\begin{tabular}{cc}
\epsfig{file=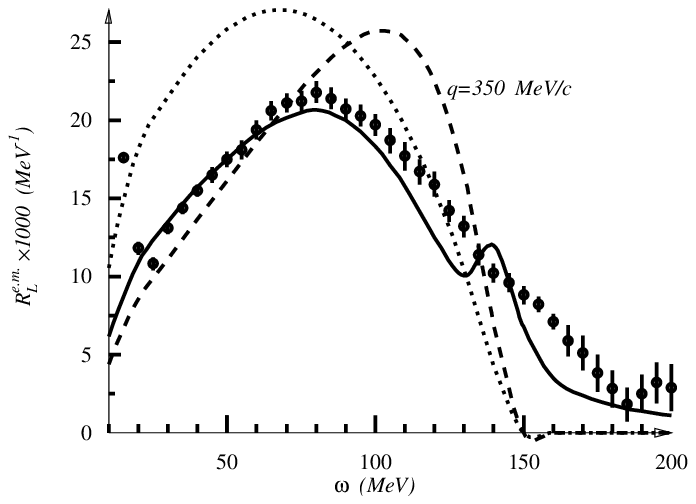}
&
\epsfig{file=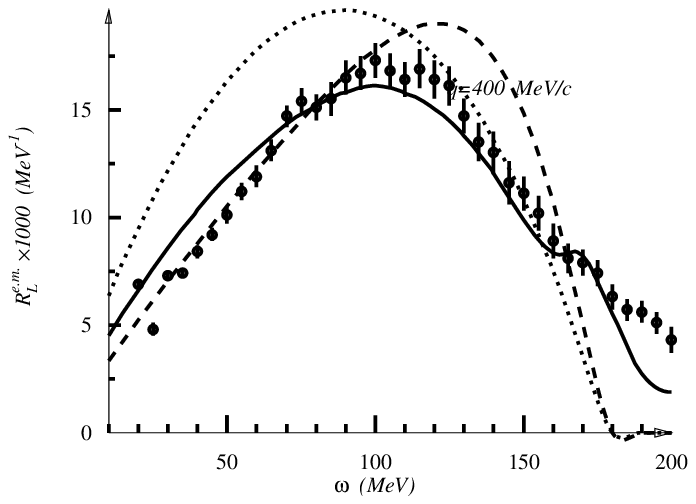}
\cr
\epsfig{file=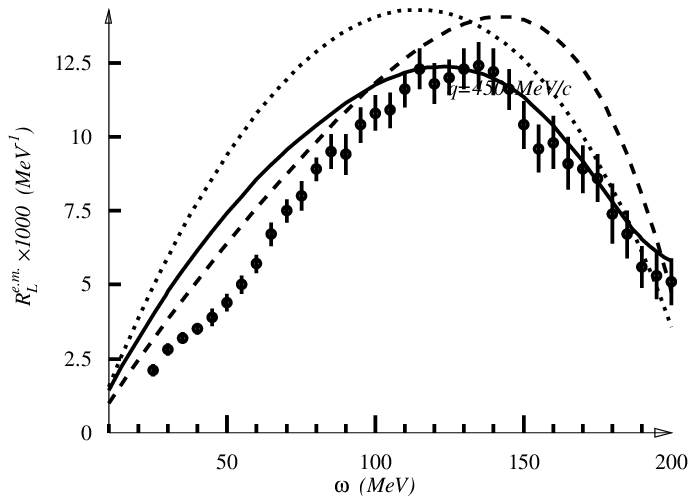}
&
\epsfig{file=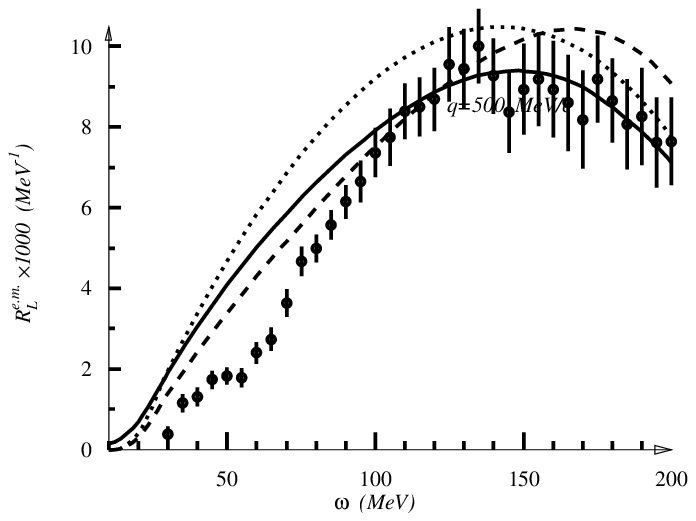}
\end{tabular}
}
\end{center}
\caption{
$R_L(q,\omega)$ for $^{12}C$ at $q~=~350,~400,~450,~500$~MeV/c.
Data from Saclay.
}
\label{fig:CL}
\end{figure}

\begin{figure}
\begin{center}
\mbox{
\begin{tabular}{cc}
\epsfig{file=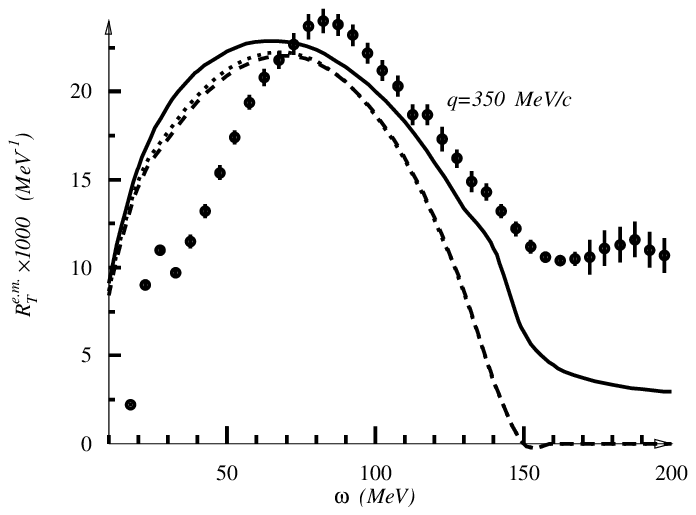}
&
\epsfig{file=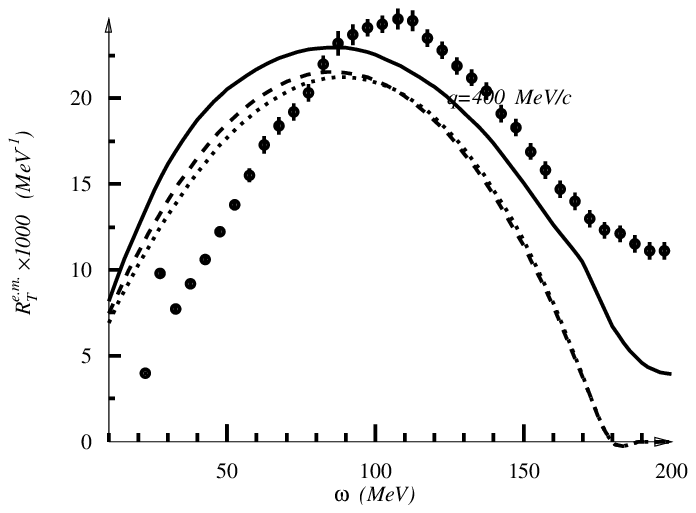}
\cr
\epsfig{file=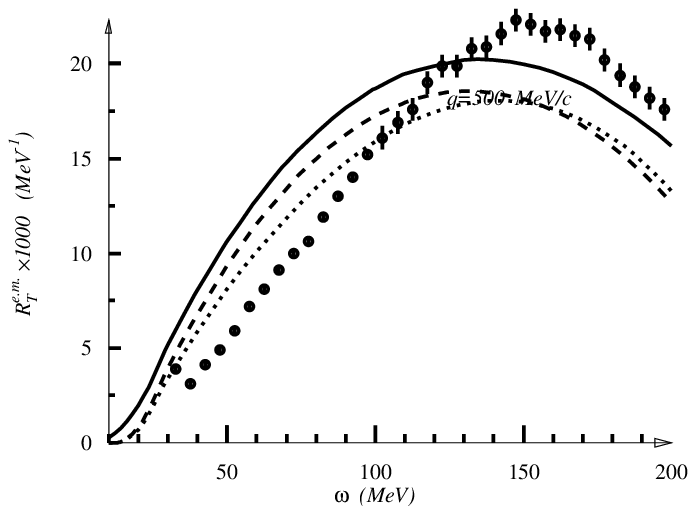}
&
\epsfig{file=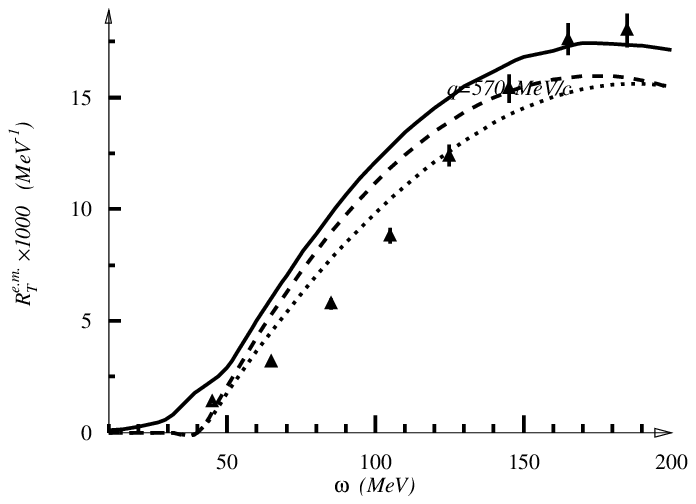}
\end{tabular}
}
\end{center}
\caption{
$R_T(q,\omega)$ for $^{12}C$ at $q~=~350,~400,~500,~570$~MeV/c.
Dots: data from Saclay, triangles: world data
}
\label{fig:CT}
\end{figure}

\begin{figure}
\begin{center}
\mbox{
\begin{tabular}{cc}
\epsfig{file=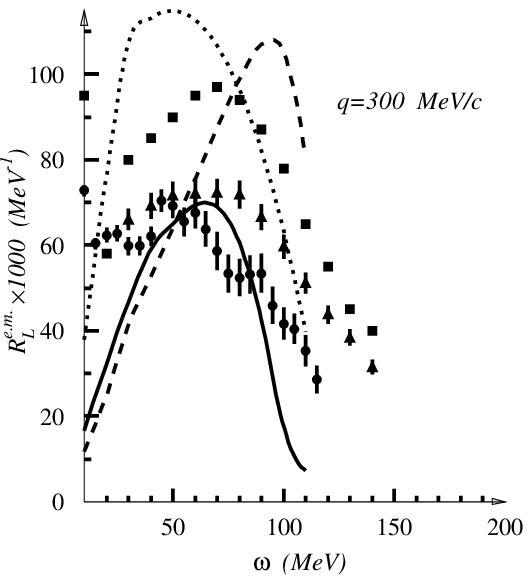}
&
\epsfig{file=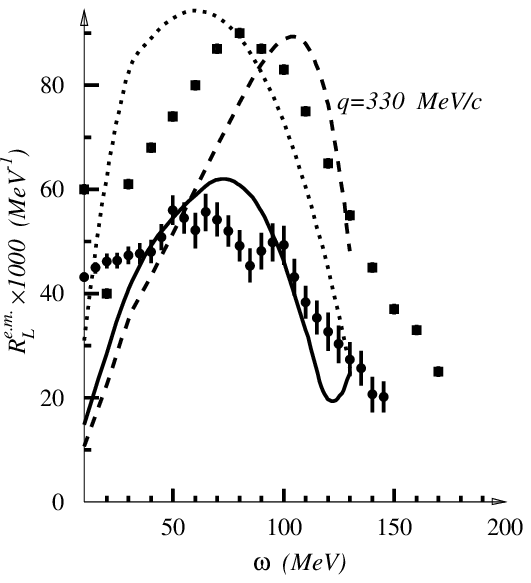}
\cr
\epsfig{file=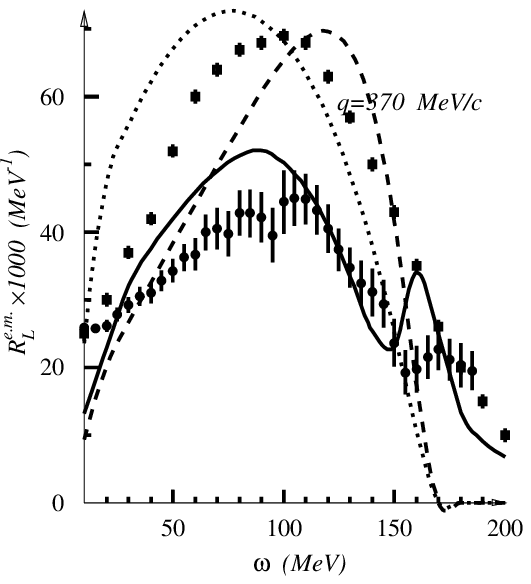}
&
\epsfig{file=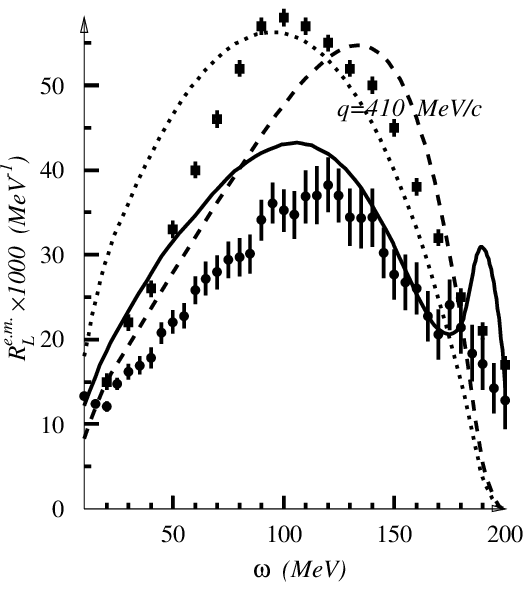}
\end{tabular}
}
\end{center}
\caption{
$R_L(q,\omega)$ for $^{40}Ca$ at $q~=~300,~330,~370,~410$~MeV/c
Dots: data from Saclay, squares: data from Bates, triangles: world data
}
\label{fig:CaL}
\end{figure}

\begin{figure}

\begin{center}
\mbox{
\begin{tabular}{cc}
\epsfig{file=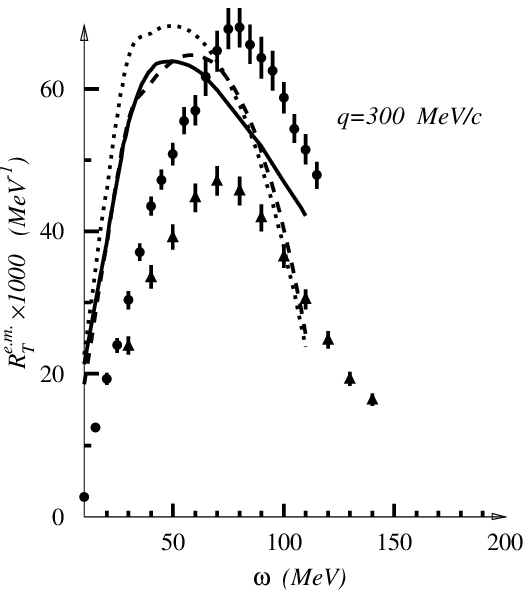}
&
\epsfig{file=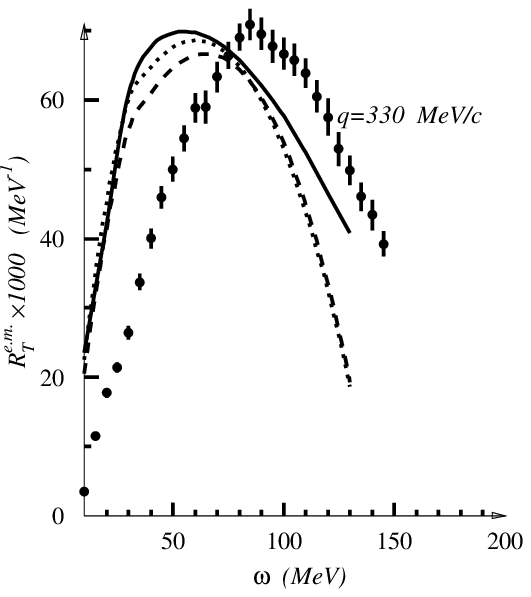}
\cr
\epsfig{file=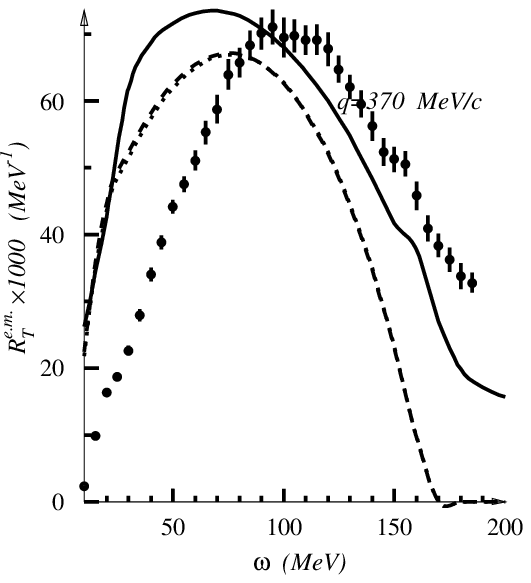}
&
\epsfig{file=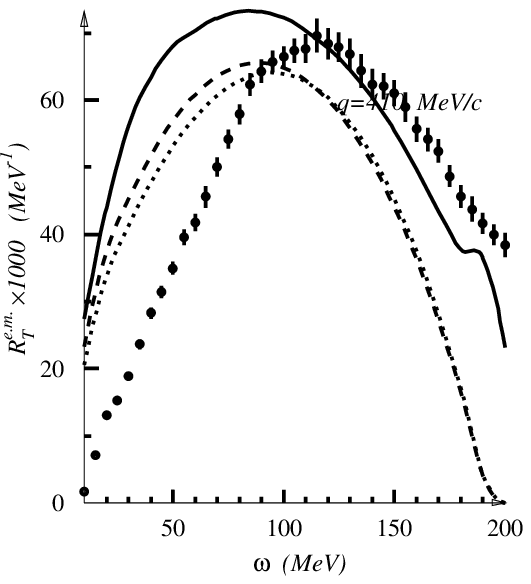}
\end{tabular}
}
\end{center}
\caption{
$R_L(q,\omega)$ for $^{40}Ca$ at $q~=~300,~330,~370,~410$~MeV/c
Dots: data from Saclay, triangles: world data
}
\label{fig:CaT}
\end{figure}

With these ingredients we obtain the results shown in Figs. \ref{fig:CL},
\ref{fig:CT}, \ref{fig:CaL} and \ref{fig:CaT}. In all these plots solid
lines are the full calculation, dashed lines the 0-loop result, while
dotted lines represent FFG outcomes. As one can see there is generally
a good overall agreement between our calculation and the known experimental
data from the Saclay experiments, which we choose as reference. In particular
$R_L$ is described quite accurately both for $^{12}C$ and $^{40}Ca$ for 
all the momenta examined.

Clearly agreement turns out to be better at higher momenta, where the
convergence of the loop expansion is expected to be faster, while some
problems begin to be evident at the lowest value of q=300~MeV.

\begin{figure}
\begin{center}
\epsfig{file=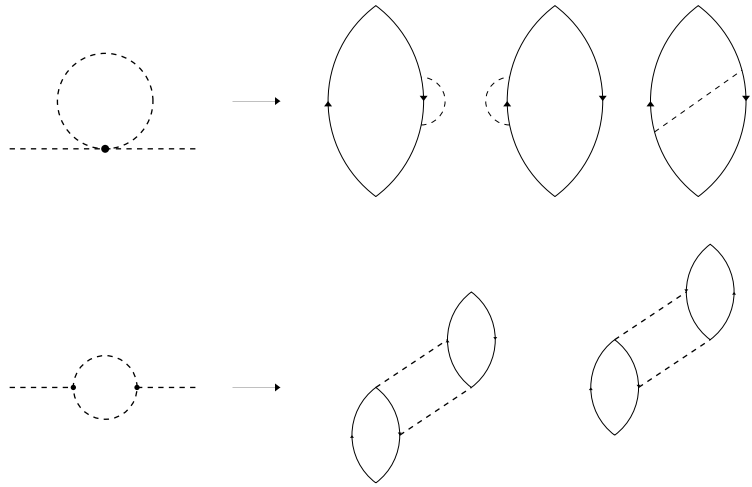}
\end{center}
\caption{The class of diagrams to be evaluated at the 1-loop order to
obtain Response functions.}
\label{fig:diag}
\end{figure}

The shape and form of $R_T$ are also described fairly well, even if 
the QEP turns out to be too much broad for $^{12}C$. The position
of the peak for $R_T$ is shifted of tens of MeV, a fact that
could be linked to a known bias of our approach. In fact, as one can see
from Fig.~\ref{fig:diag}, we cannot sum the Dyson
series for the nucleonic self-energies, at least at any given order
in the loop-expansion, a fact which conversely prevent us from the
possibility of obtaining any shift in the QEP position.

\section{Conclusion and outlook}
We have shown how a 1-loop calculation in a BLE with a reasonable choice
for the effective interaction is able to explain the disagreement
between $R_L$ and $R_T$ and the FFG predictions. 
The apparent discrepancy in the behaviour
between experiments on different nuclei can be understood in this frame
because the various contributions to the responses depend differently
upon the density of the system - in particular the diagrams of the first line
in are proportional to the density while the others to its square. Some residual
problems with experimental data, even after Jourdan's work, suggest
a renewed experimental interest into the topic. In particular world data
seems to be too much lowered for $R_T$ on $^{40}Ca$ with respect to the
original data, a feature emerged also in other calculations\cite{Fa-96}.

\begin{center}
\begin{tabular}{|c|c|c|c|c|c|c|c|}
\hline
 S & T & $C_{NN}$ & $C_{N\Delta}$ & $C_{\Delta\Delta}$ &
$\Lambda_{NN}$ & $\Lambda_{N\Delta}$ & $\Lambda_{\Delta\Delta}$ \cr
\hline
 0 & 0 & .15    & - & .15    & 1000 & - & 1000  \cr
 L & 0 & -      & - & -      & -    & - & -     \cr
 T & 0 & 1.5    & - & 1.5    & 1000 & - & 1000  \cr
 0 & 1 & .00436 & - & .00436 & 1000 & - & 1000  \cr
 L & 1 & .08    & .32 & .016 & 1300 & 1000 & 1000  \cr
 T & 1 & 2.3    & 2.3 & 2.3  & 1750 & 1000 & 1000  \cr
\hline
\end{tabular}
\end{center}
\centerline{Pion channel values are expressed as $f^2_{\pi xx}/4\pi$}

\newpage

\newpage

\newpage

\newpage

\newpage

\end{document}